# Universal quantum computing with superconducting charge qubits

Nian-Quan Jiang,[1,2*] Yao Chen,[1] Chuanbing Cai,[3] Ming-FengWang,[1Δ] Junwang Tang[2#]

[1]College of Mathematics, Physics and Electronic Information Engineering, Wenzhou University, Wenzhou 325035, China.
[2]Department of Chemical Engineering, UCL,Torrington Place, WC1E 7JE, London, United Kingdom.
[3]Research Center for Superconductors and Applied Technologies, Physics Department, Shanghai University, Shanghai, China

Superconducting quantum circuit is a promising system for building quantum computer. With this system we demonstrate the universal quantum computations, including the preparing of initial states, the single-qubit operations, the two-qubit universal gate operations between arbitrary qubits, the multiple pairs of two-qubit gate operations in parallel, the coupling operations on a group of qubits in parallel, the coupling operations on multiple groups of qubits in parallel, the coupling operations on multiple pairs and multiple groups of qubits in parallel. Within available technology, a universal quantum computer consists of more than 50 qubits allowing $10^3$ operations is achievable with the design.

Quantum computers perform computations by manipulating quantum bits and quantum entangled states, which obey the laws of quantum mechanics instead of classical physics. Therefore, a quantum computer could execute a multitude of parallel tasks exponentially faster than a classical one in certain applications such as prime factoring and quantum simulation [1-5]. In the past two decades, the researches on exploring quantum computers have attracted much attentions and tremendous progresses have been made with various systems such as ionic [6-9], photonic [10-14], superconducting [15-19], and solid-state [20-24]. For example, high-fidelity one and two qubit gate operations at the fault tolerant threshold for the surface code and high-fidelity multi-qubit entanglements have already been achieved with nearest-neighbour capacitive coupling [25]. But the hurdles encountered in realizing quantum devices are enormous, and it is still difficult to create a universal quantum computer (UQC) because challenges such as performing high fidelity operations between and among arbitrary qubits.

A UQC requires to couple arbitrary qubits, so any qubit in a computer should interact with all the others. There are two ways to implement this type of interaction: all the qubits are directly connected with each other or indirectly connected through intermediaries. A presented approach for indirectly connecting multiple qubits is known as "quantum bus", in the system of superconducting Josephson junctions, it can be implemented with various intermediaries such as LC-oscillator [26], transmission-line-rasonator [27], etc. However, in each of the existing "bus" schemes, there are two types of coupling simultaneously, i.e., coupling directly with capacitor (or inductor) and coupling indirectly through the "bus" [28]. They act on the qubits at the same time, then the unwanted couplings make it difficult to perform universal computations in a system with a large number of qubits. For a variety of reasons such as mentioned above, a practical UQC is still far away from us, and the creating of a UQC will require further appreciable advances, both practical and conceptual, in all aspects of Hamiltonian design and control. Here, we demonstrate a simple but effective scheme to implement universal quantum computing by coupling all qubits directly with capacitors, the coupling can be switched on and off by tuning the controlling parameters.

We now show that various operations needed for building a UQC can be achieved by directly capacitive coupling all the qubits and adjusting the gate voltages, external magnetic fluxes and microwave pulses. Our scheme is implemented with the architecture in Fig. 1, which consists of

$N$ superconducting charge qubits (SQUIDs, Transmons [29], or Xmons [30] ) interconnected directly with capacitors (here, we take advantage of the fact that two capacitors $C_i$ and $C_j$ connected in series is equivalent to one capacitor $C$ when their capacitances meet the condition $C^{-1} = C_i^{-1} + C_j^{-1}$ and the inductance between them is small and can be neglected). Each qubit $Q_i$ can be controlled by gate voltage $V_{gi}$, flux $\Phi_{ei}$ and microwave (XY) pulses [25] and can be measured with a readout resonator R [ 31, 25 ]. The circuit dynamics is governed by the Hamiltonian ( the constant terms are omitted ):

$$H = \sum_{i=1}^{N} E_{ci} \left( \hat{n}_i - n_{gi} \right)^2 + \sum_{i<j;\ i,j=1}^{N} E_{ij} \left( \hat{n}_i - n_{gi} \right)\left( \hat{n}_j - n_{gj} \right) - \sum_{i=1}^{N} E_{Ji} \cos \varphi_i \ . \tag{1}$$

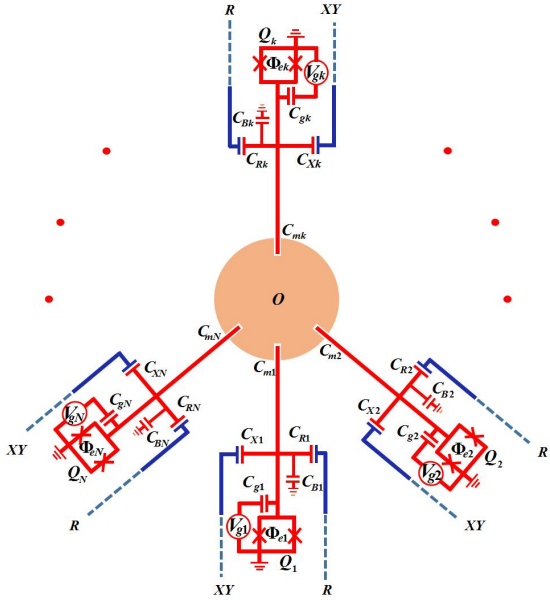

Fig. 1. Design of a universal quantum computer. Each Xmon qubit $Q_i$ is connected to the superconducting metal body (orange area denoted as "o") by a capacitor $C_{mi}$, the inductance between arbitrary two capacitors $C_{mi}$ and $C_{mj}$ is very small and can be neglected. Thus, arbitrary two qubits $Q_i$ and $Q_j$ are directly coupled with capacitor with the value of capacitance $C_{mi}C_{mj}/(C_{mi}+C_{mj})$. The effective Josephson energy $E_{Ji}$ of each qubit $Q_i$ is controlled by the external magnetic flux $\Phi_{ei}$ through the superconducting loop of the qubit.

Single-qubit operations can be performed using pulses on the microwave XY control line (blue), which is connected to the island of qubit $Q_i$ with a coupling capacitor $C_{Xi}$. The gate voltage given by the XY control line is $V_{Xi}$. Each qubit can be measured by using a coplanar waveguide resonator R (blue), the capacitance between its input port and the island is $C_{Ri}$ and the gate voltage given by the resonator is $V_{Ri}$. A shunted capacitor $C_{Bi}$ is used to ensure $C_{mi} + C_{Bi} \gg C_i$ so that the qubit is an Xmon. The coupling between arbitrary two qubits can be turned on or off by changing the fluxes so that $\omega_i = \omega_j$, $\omega_i + \omega_j \gg J_{ij}$ or $|\omega_i - \omega_j| \gg J_{ij}$, respectively.

Where $E_{ci} = 2e^2 \left(1 + C_{mi}^2 / \beta D_i \right) / D_i$ and $E_{ij} = 4e^2 C_{mi} C_{mj} / \beta D_i D_j$ ($i \neq j$, $i, j = 1, 2, \cdots, N$) are the effective Cooper-pair charging energy of qubit $i$ and the coupling energy between qubits $i$ and $j$, respectively. $e$ is the charge on the electron. $D_i = C_i + C_{gi} + C_{mi} + C_{Bi} + C_{Ri} + C_{Xi}$ is the sum of all capacitances connected to the island of qubit $i$ including Josephson capacitance $C_i$ and gate capacitance $C_{gi}$. $\beta$ is related to all the capacitances with $\beta = \sum_{i=1}^{N} C_{mi} (D_i - C_{mi}) / D_i$. $\hat{n}_i$ is the number of excess Cooper pairs in the island. $n_{gi} = -(C_{gi} V_{gi} + C_{Ri} V_{Ri} + C_{Xi} V_{Xi}) / 2e$ is the normalized charges induced on the qubit by gate voltages. $\varphi_i$ is the phase difference across the junctions of the qubit (assume the two junctions of the qubit are identical). The effective Josephson coupling energy $E_{Ji}$ is tunable by the external flux $\Phi_{ei}$ between $2E_{Ji}^0$ and zero:

$$E_{Ji} = 2E_{Ji}^0 \cos\left(\pi \Phi_{ei} / \Phi_0\right), \tag{2}$$

where $E_{Ji}^0$ is the Josephson coupling energy of single junction in the qubit $i$, $\Phi_0 = h/2e$ is the flux quantum.

Truncate the Hamiltonian (1) to the subspace spanned by the eigen states of $\hat{n}_i$, i.e., $|0_i\rangle$ and $|1_i\rangle$, one can derive $\cos\varphi_i = \sigma'_{xi}/2$ with $\sigma'_{xi} = |1_i\rangle\langle 0_i| + |0_i\rangle\langle 1_i|$. The eigen states of $\sigma'_{xi}$ are denoted as $|\pm_i\rangle = (|0_i\rangle \pm |1_i\rangle)/\sqrt{2}$. In the basis of $|\pm_i\rangle$ and at the charge degeneracy point where $n_{gi} = 1/2$, the Hamiltonian (1) reduces to

$$H = -\sum_{i=1}^{N} \frac{\hbar\omega_i}{2}\sigma_i^z + \sum_{i<j;\ i,j=1}^{N} \hbar J_{ij}\sigma_i^x \sigma_j^x, \qquad (3)$$

where $\omega_i = E_{Ji}/\hbar$, $J_{ij} = E_{ij}/4\hbar$, and the Pauli matrices $\sigma_i^z = |+_i\rangle\langle +_i| - |-_i\rangle\langle -_i|$ and $\sigma_i^x = |+_i\rangle\langle -_i| + |-_i\rangle\langle +_i|$. When $|\omega_i - \omega_j| \gg J_{ij}$, in the rotating-wave approximation (RWA) the interaction term $\sigma_i^x \sigma_j^x$ in the Hamiltonian (3) can be neglected. On the other hand, when $\omega_i = \omega_j$ and $\omega_i + \omega_j \gg J_{ij}$, in the RWA the term $\sigma_i^x \sigma_j^x$ will be equal to $\sigma_i^+ \sigma_j^- + \sigma_i^- \sigma_j^+$ with $\sigma_j^\pm = (\sigma_j^x \pm i\sigma_j^y)/2$.

In the following, based on the theories mentioned above we demonstrate various operations needed for performing universal quantum computations. Firstly, the system has to be prepared in an initial state. For this we keep the gate voltages of all the qubits in the system at degeneracy point, i.e., $n_{gi} = 1/2$, and tune the fluxes $\Phi_{ei}$ at low temperature to get $\hbar\omega_i \gg k_B T$ and $|\omega_i - \omega_j| \gg J_{ij}$ ($i \neq j$; $i,j = 1,2,\cdots,N$). When the evolving time $t \gg \pi/|\omega_i - \omega_j|$, in the RWA the Hamiltonian (3) reads $H = -\sum_{i=1}^{N} \hbar\omega_i \sigma_{zi}/2$. After sufficient time, the residual interaction relaxes all the qubits in the system to the ground states, i.e., the system will be prepared in the initial state $|+_1 +_2 \cdots +_N\rangle$.

Then the single-qubit gate operations have to be performed. Using pulses on the microwave (XY) line in Fig. 1, rotations around the X and Y axes in the Bloch sphere representation can be performed [25, 31]. To operate single qubit $i$ we set the gate voltages of all the qubits at degeneracy point $n_{gl} = 1/2$ ($l = 1, 2, \cdots, N$), while tune the fluxes to satisfy $|\omega_i - \omega_j| \gg J_{ij}$ and $|\omega_j - \omega_k| \gg J_{jk}$, where $j, k \neq i$; $j \neq k$; $j, k = 1, 2, \cdots, N$. Then single-qubit gates on qubit $i$ can be performed using microwave pluses in the way similar to that taken in the reference [25].

A two-qubit gate operation on arbitrary qubits $i$ and $j$ can be achieved by controlling the corresponding gate voltages and magnetic fluxes. Tune the fluxes $\Phi_{ei}$ and $\Phi_{ej}$ to get $\omega_i = \omega_j$ and $|\omega_i + \omega_j| \gg J_{ij}$, while the fluxes of all other qubits are properly tuned to satisfy the conditions $|\omega_j - \omega_k| \gg J_{jk}$, $|\omega_i - \omega_k| \gg J_{ik}$ and $|\omega_l - \omega_m| \gg J_{lm}$, where $k, l, m \neq i, j$; $k, l, m \in \{1, 2, \cdots, N\}$, and the gate voltages of all the qubits in the system are at degeneracy point. In the RWA, the Hamiltonian of the system reduces to

$$H = -\sum_{i=1}^{N} \frac{\hbar \omega_i}{2} \sigma_i^z + \hbar J_{ij} \left( \sigma_i^+ \sigma_j^- + \sigma_i^- \sigma_j^+ \right). \tag{4}$$

After a period of evolving time $\tau = \pi / (2 J_{ij})$, the Hamiltonian (4) will produce a swapping operation: $|+_i -_j\rangle \rightleftarrows |-_i +_j\rangle$. When the evolving time $\tau = \pi / (4 J_{ij})$, the Hamiltonian (4) should correspond to a $\sqrt{i\text{SWAP}}$, which is a universal two qubit gate for the states $|-_i -_j\rangle$, $|-_i +_j\rangle$, $|+_i -_j\rangle$, $|+_i +_j\rangle$. After that, $\Phi_{ei}$ and $\Phi_{ej}$ are set back to satisfy $|\omega_i - \omega_j| \gg J_{ij}$ and then the coupling between qubits $i$ and $j$ is turned off.

Performing multiple pairs of two-qubit universal gate operations in Parallel can be achieved as follows. If we want to simultaneously perform two-qubit universal gates on $k$ pairs of qubits: $i_1$ and $j_1$, $i_2$ and $j_2$, ......, and $i_k$ and $j_k$, respectively. We tune the magnetic fluxes $\Phi_{ei_m}$ and $\Phi_{ej_m}$, ($i_m \neq j_m$; $i_m, j_m \in \{1, 2, \cdots, N\}$; $m = 1, 2, \cdots, k$), so that $\omega_{i_m} = \omega_{j_m}$, $\omega_{i_m} + \omega_{j_m} \gg J_{i_m j_m}$ and $|\omega_{i_m} - \omega_{i_{m'}}| \gg J_{i_m i_{m'}}$, ($m \neq m'$; $m, m' = 1, 2, \cdots, k$), while the fluxes of all other qubits are tuned to satisfy $|\omega_n - \omega_l| \gg J_{nl}$, $|\omega_{i_m} - \omega_l| \gg J_{i_m l}$ and $|\omega_{j_m} - \omega_l| \gg J_{j_m l}$, where $n \neq l$; $n,l \neq i_m, j_m$; $m = 1, 2, \cdots, k$; $n, l \in \{1, 2, \cdots, N\}$, and the gate voltages of all the qubits in the system are at degeneracy point. In the RWA, the Hamiltonian of the system is described by

$$H = -\sum_{k=1}^{N} \frac{\hbar \omega_k}{2} \sigma_k^z + \sum_{m=1}^{k} \hbar J_{i_m j_m} \left( \sigma_{i_m}^+ \sigma_{j_m}^- + \sigma_{i_m}^- \sigma_{j_m}^+ \right). \qquad (5)$$

After a period of evolving time $\tau = \pi / (2 J_{i_m j_m})$, for convenience, assume $J_{i_m j_m} \equiv J$, ($m = 1, 2, \cdots, k$), the Hamiltonian (5) will produce $k$ pairs of swapping operations: $\left| +_{i_m} -_{j_m} \right\rangle \rightleftarrows \left| -_{i_m} +_{j_m} \right\rangle$, ($m = 1, 2, \cdots, k$). When evolving time $\tau = \pi / (4 J_{i_m j_m})$, the Hamiltonian (5) should correspond to $k$ universal two qubit $\sqrt{i\text{SWAP}}$ gates. These gates operate on $k$ pairs of qubits in parallel, which leads to that $k$ pairs of qubits are entangled respectively, but the qubits come from different pairs are not coupled. After that, the fluxes are set back to satisfy $\left| \omega_{i_m} - \omega_{j_m} \right| \gg J_{i_m j_m}$ and then the interactions are turned off.

To perform coupling operations on a group of qubits (more than two qubits, say qubits 1 through $k$, $k > 2$), we can control the magnetic fluxes of the selected qubits to make their effective Josephson energies the same, so that $\omega_i = \omega_0$ and $2\omega_0 \gg J_{ij}$, ($i \neq j$; $i, j = 1, 2, \cdots, k$), while the fluxes of all other qubits are tuned to satisfy $\left| \omega_l - \omega_m \right| \gg J_{lm}$ and $\left| \omega_l - \omega_i \right| \gg J_{li}$ with $l \neq m$; $l, m = k+1, k+2, \cdots, N$; $i = 1, 2, \cdots, k$, and the gate voltages of all the qubits in the system are at degeneracy point. In the RWA, the Hamiltonian of the system is governed by

$$H = -\sum_{i=1}^{N} \frac{\hbar \omega_i}{2} \sigma_i^z + \sum_{i<j;\ i,j=1}^{k} \hbar J_{ij} \left( \sigma_i^+ \sigma_j^- + \sigma_i^- \sigma_j^+ \right). \qquad (6)$$

After a period of evolving time $\tau = \pi / (4 J_{ij})$, for convenience, assume $J_{ij} \equiv J_{12}$ ($i \neq j, i, j = 1, 2, \cdots, k$), the Hamiltonian (6) should correspond to a series of $\sqrt{i\text{SWAP}}$, which operate in parallel on arbitrary pair of qubits $i$ and $j$ ($i \neq j$; $i,j = 1,2,\cdots,k$) in the group, thus in the group any qubit is entangled with all the others. After that, the fluxes are set back to satisfy $|\omega_i - \omega_j| \gg J_{ij}$ and then the interactions are turned off.

Coupling operations on multiple groups of qubits in parallel are also achievable with the presented architecture. Assume $l$ groups of qubits are selected, there are $k_i$ qubits in group $i$ ($i = 1,2,\cdots,l$). Simultaneously tune the magnetic fluxes of all selected qubits in the same group to make their effective Josephson energies are the same, but they are different for the qubits coming from different groups and the differences of the effective Josephson energies for the arbitrary two qubits in different groups are much larger than their coupling energies, i.e., $\omega_{i_m} = \omega_{i_0}$, $2\omega_{i_0} \gg J_{i_m i_{m'}}$ and $|\omega_{i_0} - \omega_{j_0}| \gg J_{i_m j_n}$, where $m \neq m'$; $i_m \neq i_{m'}$; $m, m' = 1,2,\cdots,k_i$; $i \neq j$; $i,j = 1,2,\cdots,l$; $n = 1,2,\cdots,k_j$; $i_m \neq j_n$; $i_m, i_{m'}, j_n \in \{1,2,\cdots,N\}$. While the fluxes of all other qubits are tuned to satisfy $|\omega_{i_m} - \omega_r| \gg J_{i_m r}$ and $|\omega_s - \omega_r| \gg J_{sr}$, here $i_m, r, s \in \{1,2,\cdots,N\}$; $r \neq s$; $r, s \neq i_m$; $i = 1,2,\cdots,l$; $m = 1,2,\cdots,k_i$, and the gate voltages of all the qubits in the system are at degeneracy point. In the RWA, the Hamiltonian of the system is governed by

$$H = -\sum_{i=1}^{N} \frac{\hbar \omega_i}{2} \sigma_i^z + \sum_{i=1}^{l} \sum_{m<m'; \, m,m'=1}^{k_i} \hbar J_{i_m i_{m'}} \left( \sigma_{i_m}^+ \sigma_{i_{m'}}^- + \sigma_{i_m}^- \sigma_{i_{m'}}^+ \right) \quad (7)$$

After the evolving time $\tau = \pi/(4J_{i_m, i_{m'}})$, the qubits $i_m$ and $i_{m'}$ in the group $i$ are coupled by a $\sqrt{i\text{SWAP}}$. In the case of $J_{i_m i_{m'}} = J$ ($m \neq m'$, $m,m' = 1,2,\cdots,k_i$, $i = 1,2,\cdots,l$), after the evolving time $\tau = \pi/(4J)$, arbitrary two qubits in the same group are simultaneously coupled by a $\sqrt{i\text{SWAP}}$ gate operation and all $l$ groups are evolving in parallel. But the Hamiltonian (7) will not cause coupling for the qubits coming from different groups. After that, the system is set back to the idle state where the interactions are turned off, .

Coupling multiple pairs of qubits and multiple groups of qubits in Parallel can also be performed. If we want to simultaneously perform two-qubit universal gates on $k$ pairs of qubits (i.e., the qubits $i_m$ and $j_m$, $m=1,2,\cdots,k$, $i_m, j_m \in \{1,2,\cdots,N\}$) and coupling operations on $k'$ groups of qubits (i.e., the qubits $r^n$, $r=1,2,\cdots,l_n$, $n=1,2,\cdots,k'$, $r^n \in \{1,2,\cdots,N\}$, there are $l_n$ qubits in the group $n$), we tune the magnetic fluxes so that the frequencies of the qubits in any pair satisfy $\omega_{i_m} = \omega_{j_m}$, $\omega_{i_m} + \omega_{j_m} \gg J_{i_m j_m}$ with $i_m \neq j_m$, but the differences of the frequencies between any two qubits coming from different pairs are much large than their coupling strength, while the frequencies of qubits in any group satisfy $\omega_{r^n} = \omega_{r^{n'}}$, $\omega_{r^n} + \omega_{r^{n'}} \gg J_{r^n r^{n'}}$ with $n \neq n'$ and $n, n' = 1,2,\cdots,k'$, but the differences of the frequencies between any two qubits coming from different groups are much large than their coupling strength. Meanwhile, for any two qubits, one from the $k$ pairs or the $k'$ groups and the other from the qubits outside the $k$ pairs and the $k'$ groups, the differences of the frequencies between them are much large than their coupling strength. The differences of the frequencies between any two qubits outside the $k$ pairs and the $k'$ groups are much large than their coupling strength. The gate voltages of all the qubits in the system are at degeneracy point. In the RWA, the Hamiltonian of the system is

$$H = -\sum_{i=1}^{k} \frac{\hbar \omega_i}{2} \sigma_i^z + \sum_{m=1}^{k} \hbar J_{i_m j_m} \left( \sigma_{i_m}^+ \sigma_{j_m}^- + \sigma_{i_m}^- \sigma_{j_m}^+ \right) + \sum_{n<n'; n,n'=1}^{k'} \sum_{r=1}^{l_n} \hbar J_{r^n r^{n'}} \left( \sigma_{r^n}^+ \sigma_{r^{n'}}^- + \sigma_{r^n}^- \sigma_{r^{n'}}^+ \right) \quad (8)$$

After the evolving time $\tau = \pi/(4J)$ (for convenience, here we assume $J_{i_m j_m} = J_{r^n r^{n'}} = J$), each pair of the selected qubits is coupled by a $\sqrt{i\text{SWAP}}$ and any two qubits in each group of selected qubits are coupled with a $\sqrt{i\text{SWAP}}$, these coupling operations are performed in parallel. But the qubits come from different pairs or groups are not coupled. After that, the system is set back to the idle state where the interactions are turned off.

Several conditions have been assumed in the design in order to obtain a controlled manipulation of qubits. Here we discuss the appropriate range of parameters. For the transmon (or Xmon) qubits, the Josephson energy $2E_{Ji}^0/h \approx 50\text{GHz}$ is an experimentally accessible value [32], then the frequency $\omega_i/2\pi$ of charge qubits is tunable in the range of $0 \sim 50$ GHz. To ensure the Xmons are operated at a large $E_{Ji}/E_{ci}$ ratio, we can choose proper values of $C_{Bi}$ and $\Phi_{ei}$ so that $E_{Ji}/E_{ci} \geq 100$, e.g., $E_{ci}/h \sim 50\text{MHz}$ and $E_{Ji}/h \geq 5\text{GHz}$. Thus, the frequency $\omega_i/2\pi$ should be tunable in the range of $5 \sim 50$ GHz. Meanwhile, to employ the RWA, the differences of frequency between qubits have to meet the conditions $|\omega_i - \omega_j| \gg J_{ij}$, $i \neq j$. So we choose $|\omega_i - \omega_j| \sim 10^3 J_{ij}$ for the qubits which are being selected to perform operation. But, it will cause frequency crowding in a system with large number of qubits if we choose $|\omega_i - \omega_j| \sim 10^3 J_{ij}$ for all the qubits including those that will be operated after some time. To avoid this problem and simultaneously meet the conditions for employing RWA, we choose $|\omega_i - \omega_j| \sim max\{10^3 J_{ij}/k, 50 J_{ij}\}$ for the qubits that will be operated after a period of time $t = k\tau$, where $k$ is an integer and $\tau = \pi/(2J_{ij})$ is the evolving time for one coupling operation. Choose the interaction strength $J_{ij} = 10\pi$ MHz, the timescale of a coupling operation is $\tau = 50$ ns. With the above parameters, the number of qubits in the system can be chosen in the range of 50-100, and within the available coherence time $\tau_{coh} \sim 10^2 \mu s$ [29,30] we can perform operations nearly $10^3$ times. Moreover, the remaining interactions result from the RWA should be reduced as much as possible because each qubit is connected with all the others in the system and then a large number of interactions should be turned off by the RWA. For this, we employ strict conditions for the RWA mentioned above, on the other hand we can properly choose fluxes so that in the interaction picture the phases of the remaining interaction terms are distributed randomly or evenly, then the influences of all remaining interactions on each qubit will cancel each other out.

In summary, we have demonstrated universal quantum computations using a superconducting circuit. The incorporation of local flux control, gate voltage regulation and microwave pulse has enabled preparation of the initial state of the system, single-qubit gate operations, two-qubit gate operations between arbitrary qubits, multiple two-qubit gate operations in parallel, a group of multiple qubit couplings among arbitrary qubits, multiple groups of multiple qubit couplings in parallel, and multiple pairs of qubits and multiple groups of qubits couplings in Parallel. Within the current technology, the present architecture allows a universal quantum computer at the scale of more than 50 qubits with the number of coupling operations on the order of $10^3$.

______________________________________________

*Electronic address: jiangnq@wzu.edu.cn

Δ Electronic address: mfwang@wzu.edu.cn

#Electronic address: junwang.tang@ucl.ac.uk

**Acknowledgements:**

We thank Heng Fan, Shi-Ping Zhao, Yu-xi Liu, and yi-rong Jing for helpful discussions.